\documentclass[structabstract]{aa}  
\usepackage{natbib}
\usepackage{graphicx}
\usepackage{txfonts}
\usepackage{color}

\begin{document}
\title{Coreshine in L1506C - 
       Evidence for a primitive big-grain component or 
       indication for a turbulent core history?}

\author{
{J. Steinacker}\inst{1,2}
 \and
{C.W. Ormel}\inst{3}
 \and
{M. Andersen}\inst{1}
 \and
{A. Bacmann}\inst{1}
       }
\institute{
UJF-Grenoble 1 / CNRS-INSU, 
Institut de Planetologie et d'Astrophysique de Grenoble (IPAG) 
UMR 5274, Grenoble, F-38041, France\\
\email{stein@mpia.de, morten.andersen@obs.ujf-grenoble.fr, 
       aurore.bacmann@ujf-grenoble.fr}
\and
Max-Planck-Institut f\"ur Astronomie,
K\"onigstuhl 17, D-69117 Heidelberg, Germany
\and
Astronomy Department, University of California, 
Berkeley, CA 94720, USA\\
\email{ormel@astro.berkeley.edu}
          }
\date{Received ; accepted }

  \abstract
{
With the initial steps of the star formation process in the densest part
of the ISM still under debate, much attention is payed to the formation
and evolution of pre-stellar cores. 
The recently discovered coreshine effect can aid in exploring the
core properties and in probing the large grain population of the ISM.
}
{   
We discuss the implications of the coreshine detected from the molecular 
cloud core L1506C in the Taurus filament for the history of the core 
and the existence of a primitive ISM component of large grains becoming
visible in cores.
}
{
The coreshine surface brightness of L1506C is determined 
from IRAC Spitzer images at 3.6 $\mu$m.
We perform grain growth calculations to estimate the grain size distribution
in model cores similar in gas density, radius, and turbulent velocity
to L1506C.
Scattered light intensities at 3.6 $\mu$m are calculated for a variety of
MRN and grain growth distributions using
the DIRBE 3.5 $\mu$m all-sky map as external interstellar radiation field,
and are compared to the observed coreshine surface brightness.
}
{
For a core with the overall physical properties of L1506C, no detectable
coreshine is predicted with a size distribution following the shape and size
limits of an MRN distribution.
Extending the distribution to grain radii of about 0.65 $\mu$m allows to reproduce
the observed surface brightness level in scattered light.
Assuming the properties of L1506C to be preserved, models for the growth of grains in cores
do not yield sufficient scattered light to account for the coreshine within the
lifetime of the Taurus complex.
Only increasing the core density and the turbulence amplifies the scattered light intensity
to a level consistent with the observed coreshine brightness.
}
{
The coreshine observed from L1506C requires the presence of grains with sizes exceeding
the common MRN distribution.
The grains could be part of primitive omni-present large grain population becoming
visible in the densest part of the ISM, could grow under the turbulent dense conditions
of former cores, or in L1506C itself. In the later case, L1506C must have
passed through a period of larger density and stronger turbulence. 
This would be consistent with the surprisingly strong depletion 
usually attributed to high column densities, and with the large-scale 
outward motion of the core envelope observed today.
}


\authorrunning{Steinacker et al.}
\titlerunning{Coreshine in L1506C}
\maketitle

\section{Introduction} \label{intro}
Star formation research has identified the densest parts of molecular
clouds as the sites where stars form. 
But the entire process from forming the molecular cloud, through the 
formation of the cloud cores, its evolution while having no central 
object, to the final collapse is not understood
\citep{2007ARA&A..45..339B}.
Physically, we are facing a coupled problem of gas dynamics including
turbulent motions, gravitation, and magnetic fields
where cooling and heating of the gas and dust are influenced by the 
chemical processes, grain processing, and the external illumination.

The theory of gravo-turbulent star formation predicts that molecular cloud
 cores form at the stagnation points of the complex turbulent flow pattern 
\citep{2004RvMP...76..125M}.
But since supersonic turbulence does not create 
hydrostatic equilibrium configurations, the density structures are transient
and dynamically evolving, as the different contributions to virial equilibrium 
do not balance 
\citep{2003ApJ...585L.131V}. Therefore, a statistical approach in collecting
core properties has been a main approach to study their evolution.
For individual cores, their history has been accessed by studying their
chemical state \citep{2013A&A...551A..38P}.

The filamentary Taurus star formation region 
\citep[see, e.g.,][]{2012arXiv1206.2115Q}
is an ideal site to test these theories and to determine the properties of cores
due to its proximity and coverage of all low-mass star formation
stages.
In a recent analysis of a Taurus filament fragment L1506C based on dust emission,
and line emission of C$^{18}$O, N$_2$H$^+$, $^{13}$CO, and C$^{17}$O, 
\citet[][Pea10 hereafter]{2010A&A...512A...3P} show that the fragment
combines interesting and surprising properties.
Interpreted as a core-envelope system, the core located within 3$\times$10$^4$ au
shows low densities n(H$_2)<5\times$10$^{10}$ m$^{-3}$, but high depletion $>$30 of 
C$^{18}$O 
and low temperatures 8-10 K depending on the tracer.
Kinematic information was derived from line emission measured along a cut 
through the filament. 
The line-of-sight integrated emission was modeled with a 1D line transfer code. 
The authors argued that the line profiles show evidence for contraction of the core plus rotation
with an infall speed of about 100 m/s.
In turn, the velocity gradient of the envelope visible in the $^{13}$CO line
was found to be opposite in direction to that of the core as traced by the 
C$^{18}$O line.
Moreover, the core showed extremely low turbulence v$_{turb}$(FWHM)$<$68 m/s.
Based on the contraction of the core and its low density, the authors concluded
 that the core is about to become a pre-stellar core, but has not yet reached the
density criterion suggested by 
\citet{2008ApJ...683..238K} for a pre-stellar core.

An important piece of evidence was added to this picture by 
\citet{2010Sci...329.1622P}.
They reported on the detection of coreshine for about half of 110 investigated
cores, among them L1506C. According to the analysis
presented in 
\citet{2010A&A...511A...9S} for the core L183, the interstellar radiation
can be scattered at wavelengths around 4 $\mu$m when the core gas hosts 
dust grains which are about a factor of 4 larger than the maximal silicate grain size
of about 0.25 $\mu$m advocated by the MRN model.

Investigating Herschel observations of the entire filament harboring L1506C,
\citet{2013A&A...559A.133Y}
found that the grain opacity has to increase across the filament to fit simultaneously
the near-IR extinction and FIR Herschel emission profiles.
They interpreted this change to be a consequence of the coagulation of dust grains to form
fluffy aggregates with the grain average size being increased by a factor of 5. 

In this paper, we analyze the coreshine seen towards L1506C and discuss its
implication for the underlying grain size distribution and the history of the core.
In Sect.~2, we derive the range of coreshine surface brightness being consistent with
the IRAC 3.6 $\mu$m data of L1506C. For a model core, we calculate the scattered
light intensity for MRN-type size distributions as well as for
time-dependent distributions arising from coagulation calculations in Sect.~3. 
Comparing the derived surface brightnesses we
propose possible scenarios to account for the coreshine in Sect.~4.

\section{Determination of the coreshine surface brightness}
L1506C has been observed as part of the Taurus {\it Spitzer} Survey 
\citep[PI Padgett;][]{2010ApJS..186..259R}.
Two observations per pointing were performed across L1506C resulting in an 
average integration time per pixel of 25.6 seconds.
Point sources have been identified and masked using the sextractor software 
\citep{1996A&AS..117..393B}.
Saturated sources were not identified as point sources and are left in the images.
The image has been smoothed by a Gaussian kernel with a width of 2 pixels (2.4\arcsec).

\begin{figure}
\includegraphics[width=9cm]{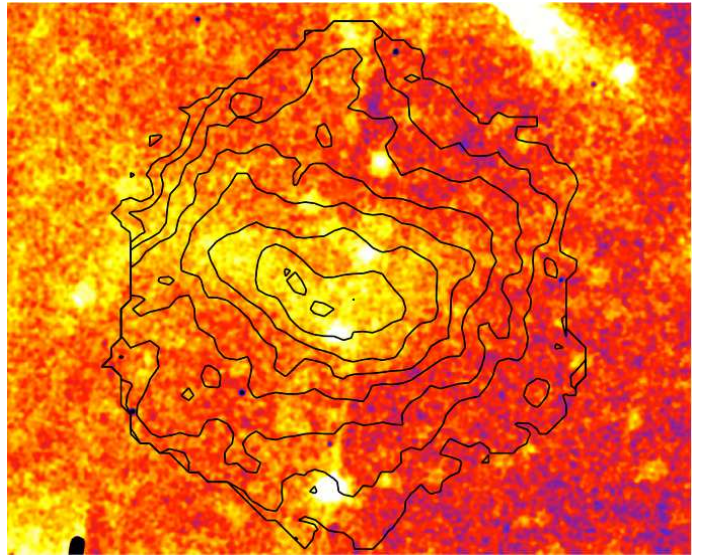}
\caption{
The core L1506C seen in coreshine and thermal dust emission.
Image: Spitzer IRAC channel 1 map of L1506C. North is up and East is to the left.
Non-saturated point sources have been identified and masked using sextractor. 
Contours: MAMBO 1.2mm observations (Pea10), in steps of 0.5 mJy/beam up to 5.5 mJy/beam. 
The field of view is 20\arcmin $\times$ 13\arcmin. 
        }
\label{cs}
\end{figure}

As visible in \citet{2010Sci...329.1622P}, Fig.~\ref{cs}, for many cores, coreshine traces well the 
extinction pattern visible in the 8 $\mu$m Spitzer image. This helps to 
identify if the scattered light indeed comes from the central core part.
But for L1506C, no extinction pattern is visible above noise at 8 $\mu$m.

However, L1506C has been observed at 1.2 mm with MAMBO II with a beam size of
11\arcsec (Pea10). Fig.~\ref{cs} shows the IRAC band 1 image together with the MAMBO II map
as contours.
The coreshine is remarkably well-correlated with the thermal emission of the central
core.

The coreshine surface brightness has been measured over a 10" square region 
at the peak of the emission associated with the core. 
The background level has been estimated for a similar size region to the west 
of the core in a region with no apparent point sources. 
The surface brightness including background was determined as 0.25$\pm$0.02 
MJy/sr, whereas the background level is 0.21$\pm$0.01 MJy/sr, 
leading to a coreshine surface brightness above background of 0.04$\pm$0.03 MJy/sr.

We also examined the cold Spitzer L1506C data in the IRAC 4.5, 5.8, and 8.0 $\mu$m bands
to derive additional constraints. All three bands do not show significant enhancement
or depression
of the surface brightness within the noise. For the 8 $\mu$m band, no
background could be derived from the DIRBE maps due to the strong difference
in the band position and width.
Therefore, we have used an upper limit of 0.03 MJy/sr (0.3 MJy/sr) for 
the 4.5 (5.8) $\mu$m surface brightness, respectively.

\section{Coreshine from a model core with various grain distributions}
Determining the scattered light surface brightness of a core requires to perform
radiative transfer calculations of the usually asymmetric core density structure
taking into account the dust opacities of the various grain species
as well as the impacting anisotropic radiation field. These quantities have
substantial uncertainty ranges, and radiative transfer is too time-consuming
to run a grid in the large parameter space with some ten parameters.
The only 3D modeling of coreshine so far was performed in 
\citet{2010A&A...511A...9S} for the core L183
for grains increasing in size with gas density.
The resulting coreshine images showed stronger gradients than observed and the
smooth observational coreshine appearance of L183 and many other
cores hints towards a grain distribution with small spatial gradients and/or an
efficient mixing of grains.
In this analysis, we therefore restrict ourselves to a simple core
density model and to determine if proposed size
distributions can account to at least reach the coreshine surface brightness level
observed in L1506C, without any attempt to model the complex shape of the core, 
and refer to future studies concerning a full 3D modeling.
Moreover, we will neglect spatial variations in the grain size distribution with reference
to the results from L183 that such variations cause stronger gradients in the
coreshine appearance than observed.

We use the spherically symmetric density distribution proposed for L1506C in Pea10,
with an outer radius of $R_c=3\times 10^4$ au, a maximal density of 
5$\times 10^{10}$ m$^{-3}$, and a kink radius of $R_c=10^4$ au where
the flat inner profile turns into a powerlaw. 
The impacting radiation field
is approximated by the Zodiacal light-subtracted 3.5 $\mu$m DIRBE map provided by 
LAMBDA\footnote{http://lambda.gsfc.nasa.gov} transformed to the core location at
the Galactic coordinates (l,b)=(171.1$^o$,-17.57$^o$).
The background radiation along the line-of-sight was derived from the DIRBE map 
since the absolute background is provided as opposed to the Spitzer observations. 
Due to the large beam of DIRBE the surface brightness in each pixel is contaminated 
by point sources and we subtracted this surface brightness using the WISE point source catalogue leading
to a value of 0.032$\pm$0.02 MJy/sr for the background.

\subsection{Models based on pristine MRN distributions}
The first scenario we discuss is to mix the core gas with
a stationary size distribution commonly found in the diffuse ISM, with a 
powerlaw distribution as described in \citet{1977ApJ...217..425M} (abbreviating
the distribution MRN) and grain radii
up to 0.25 $\mu$m. 
We use a
version of the radiative transfer code applied in \citet{2010A&A...511A...9S}
that is taking advantage of the assumption of no spatial gradients in 
the size distribution \citep[for details, see Sect.~2.4 in][]{2013arXiv1303.4998S}.
Dust opacities were taken from \citet{1984ApJ...285...89D} and
\citet{2003ApJ...598.1017D}. As shown by \cite{2011A&A...532A..43O}, 
considering more realistic porous
grains will decrease the opacities at 3.6 $\mu$m. Since the optical depth encountered
in the calculations remain below 1 in most cases, considering a filling factor of 1
means deriving an upper limit for the scattered light surface brightness.

\begin{figure}
\includegraphics[width=9cm]{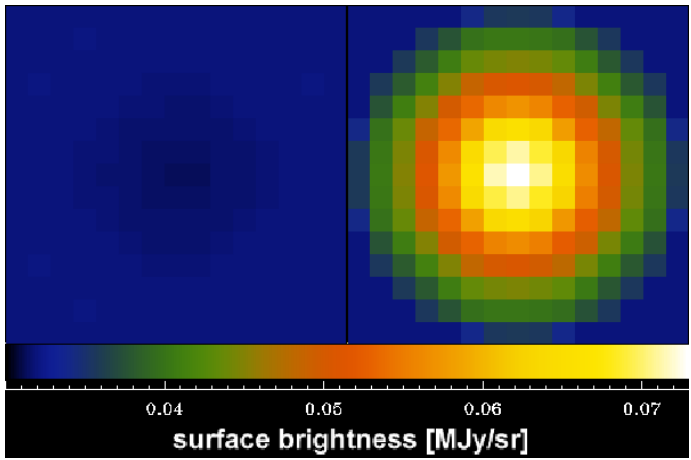}
\caption{
Scattered light images of a model core with overall properties like L1506C
using an MRN size distributions extending to grain sizes
of 0.25 $\mu$m (left) and 0.65 $\mu$m (right). The total dust mass is kept constant.
The background surface brightness is 0.032 MJy/sr and
the coreshine surface brightness above background is 0.04 MJy/sr for L1506C.
        }
\label{growth}
\end{figure}

Fig.~\ref{growth} shows images of the core at a wavelength of 3.6 $\mu$m 
for an MRN grain size distribution. 
In the left image, the distribution includes grains with
a radius up to 0.25 $\mu$m. The core is seen in slight absorption against
the background with no visible coreshine. We successively
raised the upper size limit while keeping the total dust mass constant
with a mass ratio of gas-to-dust of 120.
The right image was calculated using grains up to 0.65 $\mu$m in size, reproducing 
the observed surface brightness of coreshine in L1506C 
around 0.07 MJy/sr.
Within the uncertainties and approximations used in the modeling this means that
the coreshine reveals a large grain component being present in all parts of L1506C.
Interpreted as a pristine component, it could originate either from large-scale dust
production and processing cycles of the ISM before the Taurus region was formed or from 
collisional growth in transient density fluctuations in the filament before L1506C came to being.

\subsection{Models based on grain growth in cores}
\begin{figure}
\includegraphics[width=9cm]{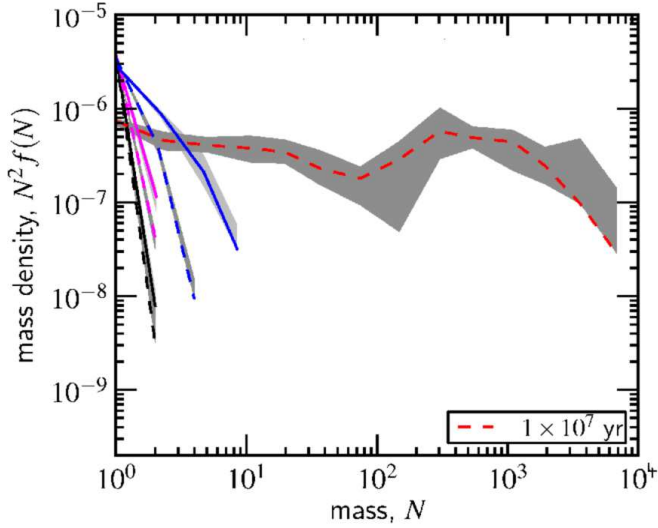}
\caption{
Mass of the size distribution per logarithmic interval as a function of the grain mass 
in units of number of initial grains.  The curves are at 
10$^4$, 3$\times$10$^4$, 10$^5$, 3$\times$10$^5$, 10$^6$, 3$\times$10$^6$, 10$^7$ yr 
continuously increasing their mass center. 
The gas density is $n=1.3\times 10^{11}$ m$^{-3}$, the turbulent speed $v_L=90$ m/s, and the core radius is $R_c=22.7$ kau.
The gray shading denotes the spread in 10 runs. The initial monomers have a size
of 0.05 $\mu$m. Only data up to 1 Myr are used in this paper.
        }
\label{O1}
\end{figure}

\begin{figure}
\includegraphics[width=9cm]{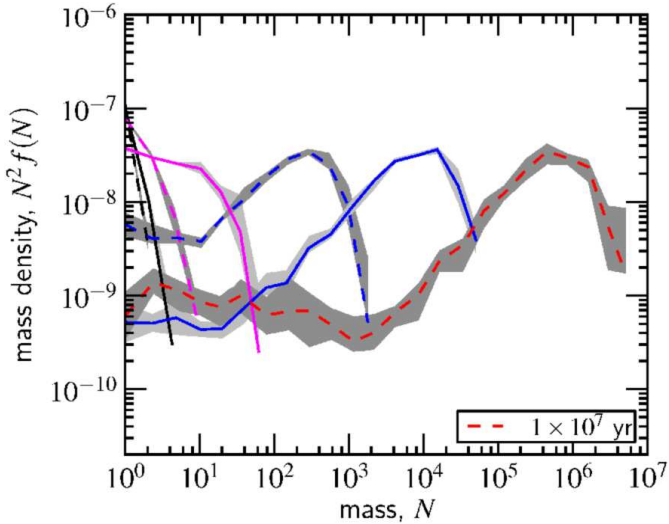}
\caption{
Same as Fig.~\ref{O1} with a different set of parameters (low density,
mean turbulent velocity):
the gas density is $n=3.8\times 10^9$ m$^{-3}$, the turbulent speed $v_L=300$ m/s, and the core radius is $R_c=22.7$ kau.
        }
\label{O2}
\end{figure}

Alternatively, we assume that the large grains are formed in the core as a consequence of
collisional growth due to the turbulent relative velocities of the grains and the 
high density. 
We utilize a time-dependent grain growth model for their distribution
which is non-trivial also because the core may change its properties during the evolution.
\citet{1993A&A...280..617O}
considered the growth of fluffy dust agglomerates in dense cores of 
molecular clouds and found that on timescales below 10$^5$ yr the
optical properties of the grains change but that no large heavy particles are
produced.
\citet{1994ApJ...430..713W} have studied the collisional evolutions of
grains in cores both in static equilibrium
and free-fall collapse. They derive powerlaw-shaped size distributions with
a cut-off at 1 $\mu$m for compact grains on a timescale of a few Myr, and
100 $\mu$m for fluffy grains. 
\citet[][hereinafter called Oea09]{2009A&A...502..845O} have improved the grain-grain
collision data by numerical experiments. 
They assumed that the decay time of the largest eddies with the size of the core is 
the sound crossing time, and that the fluctuating velocity at the largest scale is given 
by the sound speed. The turbulence on smaller scales was assumed to follow a
Kolmogorov cascade with turn-over times and velocities on the inner scale 
determined by the Reynolds number 
(for details see Sect.~2 and Appendix A in Oea09). The 
turbulence on the core scale was renewed over times larger than one free-fall time.
In their model, strongly peaked
size distributions develop as long as destructive collisions do not limit
the growth of grains, with maximal sizes also around 100 $\mu$m for a few 
free-fall times. 
Here, we apply a modified version of the Oea09 
model and have rerun a grid of coagulation calculations
for a large sample of physical
parameters extending the range compared to Oea09. 
The grid covered gas number densities from
10$^8$ to 10$^{14}$ m$^{-3}$, core radii from 4.8 to 110 kau, and
allows for fluctuating velocities at the largest scale from 90 to 510 m/s
instead of restricting to the sound speed.
The model includes destructive collisions for high relative velocities,
and in fact on longer times the grain size distributions flatten due to the
action of destructive impacts as visible in Fig.~\ref{O1} for t=$10^7$ yr.
The initial silicate grains were ice-coated spheres and had a radius of 0.05 $\mu$m. 
The grain size distributions were calculated for evolution times from 
10$^4$ to 10$^7$ yr.
Fig.~\ref{O1} and \ref{O2} show example results of calculations with parameters
partially resembling the properties of L1506C with the gas number density $n_g$,
the turbulent velocity at the outer radius $v_L$ and the core radius
$R_{cl}$. For the low turbulence case in Fig.~\ref{O1}, almost no grain growth 
is produced after 1 Myr. In the low density case (Fig.~\ref{O2}), grains sizes
around 0.7 $\mu$m were created.
Note that the results for a core similar to L1506C have been interpolated from
this grid to arrive at the correct number densities.

The Oea09 model does not consider
mixing by the same turbulence that is providing the relative velocities and
assumes the core to have a single constant density.
Since the density in our model core mimicking L1506C varies, we constructed a
global size distribution by counting for each time step the grains in each cell
grown as a function of the local density. This still overestimates the growth
at the highest densities as the mixing will at all times remove grains from 
the region towards lower density ones where growth is less efficient.
\begin{figure}
\includegraphics[width=9cm]{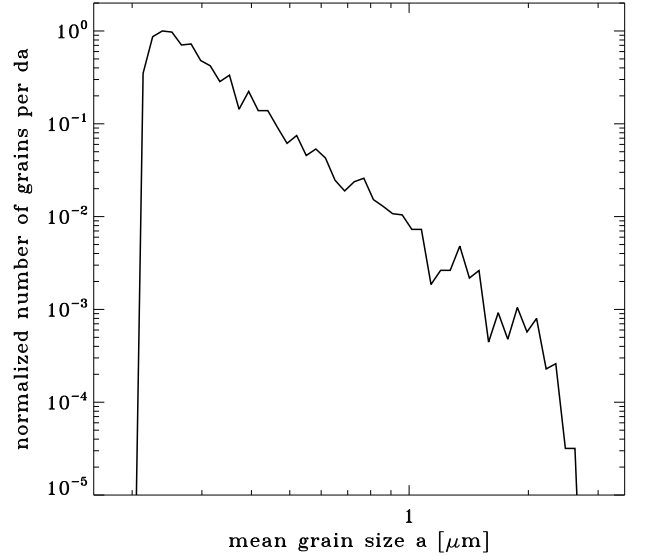}
\caption{
Normalized size distribution per size interval as a function of the
mean grain size. The distribution was obtained allowing growth at all cells
according to the local density following the Oea09 calculations and finally
mixing all grains to a constant size distribution.
        }
\label{na}
\end{figure}
The resulting size distribution per size interval is shown in Fig.~\ref{na}
for a growth time of 1 Myr and fluctuating velocities of 300 m/s.

\begin{figure}
\includegraphics[width=9cm]{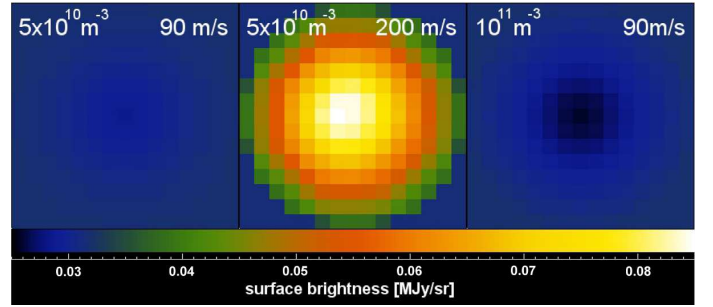}
\caption{
Scattered light images of a model core like L1506C assuming a growth time
of $10^6$ yr. In each image, the peak core density is given left,
the right number is the turbulent speed in the largest eddy.
        }
\label{growth1}
\end{figure}

We use these various size distributions to explore which of them
would create enough scattered light to explain the observed coreshine of L1506C.
We start with the currently observed turbulence and density properties.
While the free-fall time scale is of the order 10$^5$ years, thermal pressure
and magnetic fields are discussed to extend the lifetime of pre-stellar cores
from several to many free-fall times (see time scale discussion in Oea09).
Since we want to explore the conditions to produce measurable coreshine, we 
consider that the core density was at the value we observe today for at 1 Myr
in order to enable efficient grain growth.
Assuming fluctuating velocities to be 90 m/s near the observed value for the turbulent
eddy with the size of the core, {maintaining the turbulence at the largest level for 1 Myr,} the
calculation leads to no measurable coreshine surface brightness in 1 Myr (Fig.~\ref{growth1}, left).
This simple calculation suggests 
that if L1506C would have had the density distribution and the turbulence level 
that we see today with only 0.05 $\mu$m-sized grains at the onset of growth, applying
even the most optimistic growth model yields that L1506C should not 
harbor larger grains nor should it show coreshine. 

We performed radiative transfer calculations using the same set-up but
a stronger level of turbulence than currently observed.
We chose the outer eddy to have fluctuation velocities of 200 m/s
comparable to the value derived from $^{13}$CO linewidth measurements of 
about 400 m/s.
Velocities on smaller scales are smaller according to the assumed Kolmogorov
cascade. 

In the corresponding image (Fig.~\ref{growth1}, middle), coreshine is seen at the observed
surface brightness level. 
But mixing was only performed after $10^6$ yr
instead of continuous mixing. To estimate the impact
of this approximation on the largest grains, we
consider a grain that starts in the densest region. Given the centrally peaked
gas density distribution, the grain will remain in regions with a central density 
only about 1/3 ($r_0/R_c$) of the turnover time.
As coreshine is dominated by the largest grains \citep[see][]{2010A&A...511A...9S},
a more realistic mixing will result in a coreshine surface brightness too low
to account for the observed value.
Increasing the turbulent velocity even more is possible, but only up to the
sound speed since velocities in pre-stellar low-mass cores remain sub-sonic.
Moreover, the model calculations will also overestimate the growth since they assume a 
Kolmogorov turbulence decay while the linewidth measurements indicate a lack of
turbulence in the core.

Aside of considering a turbulence that has been stronger in the past, we can
also assume that the core has been denser due to the action of large-scale motions
which are tearing the core apart. This would follow the general picture of 
turbulent star formation and the higher density would let grains grow more efficiently.
The right image in Fig.~\ref{growth1} shows the core assuming the turbulence level observed today
but with a density being a factor 2 larger. Grain growth is still insufficient to 
arrive at grains large enough to produce the right coreshine level.
Finally we have chosen 300 m/s as turbulent speed of the largest eddy and
an increased density by shrinking the core and assuming the same mass.
We were able to produce coreshine well in excess of the observed value. 

We also determined the 4.5 $\mu$m surface brightness for all models described
here. The peak values never exceeded the upper limit of 0.03 MJy/sr 
derived from the data. The only exception is the model
with both increased turbulent speed and density, and in this case 
the measured 3.6 $\mu$m peak values is a tighter constraint.
In a further publication we will discuss the more recent and more
sensitive warm Spitzer measurements for this core.

\section{Conclusions} \label{conclusions}
We have investigated the scattered light observed from the central part
of L1506C in the Spitzer channel 1 and put forward two scenarios
to account for the coreshine surface brightness.
Mixing the core case with a constant MRN dust size distribution using a size
limit of 0.25 $\mu$m does not lead to sufficient scattered light to explain the
observed surface brightness of cores with coreshine.

The core may host a component of large grains with sizes up to 0.65 $\mu$m
that has not been formed in the core rather than being a primitive component
from earlier star formation cycles before the Taurus filaments came to being.
Using a simple density structure, an ISRF based on the DIRBE map plus a local component from the star formation
region,
and standard dust opacities we were able to reproduce the coreshine level with our radiative transfer
calculations.

An implication of the existence of a global primitive large-grain component would be
that most low-mass cores comparable to L1506C should show coreshine as far as 
the background level allows its detection, shielding effects do not block the radiation
either during illumination or on its way to the observer, or large-scale processes
like supernovae have modified the size distribution \citep{2011ApJ...742....7A,2012A&A...541A.154P}. A future publication
will investigate this prospective.

Another possibility is that the large grains responsible for the scattering have
been formed in the core. 
To enable efficient grain growth, we maintained the turbulence on the largest scale
for 1 Myr which is a basic assumption of the gravo-turbulent scenario for star
formation. But even with the most optimistic grain growth model,
and increasing either the turbulence or the density beyond the currently observed
values, we failed to reproduce the observed coreshine within the limits of our simple model.
However, our results indicate that grain growth in a core that is 
denser and more turbulent could lead to the observed coreshine surface brightness.
Within the gravo-turbulent scenario,
the same large-scale motions
that would have created the density maximum in the filament at the location of L1506C
could have torn the
core partially apart leaving just the gravitationally bound part.
The question is if the velocity gradients that have been inferred from the
emission line analysis along a cut in Pea10 are in agreement with this
scenario. Since turbulence on the largest scales decays slowest in the
Kolmogorov picture, it is on the large-scales that remnants of
such a tearing motion should still be seen. The observed outward motion of the
envelope at large scales can be interpreted in this way.

It is more complex to access the present kinematical signatures on the smaller scales,
that is on the scale of the core.
First, the turbulence on smaller scales decays faster and is re-created by
turbulent cascading with new kinematical signature. This leaves no trace of the history
of the fragment on small scales.
Second, the kinematic structure of the core may also be affected by self-gravity:
the change in kinematical properties appears to happen right at a Jeans length of
about 3$\times$10$^4$ au which is also the core radius. 
Furthermore, the action of magnetic
fields and density gradients can alter the behaviour of the turbulence. But the fact that
today only a low level of turbulence is observed suppressing efficient grain growth hints
towards an evolution which has seen stronger turbulence at former times.

We have not made use of the other interesting peculiarity of the core: its strong depletion.
Chemical model calculations for transient density fluctuations forming core-like density enhancements
have been performed \citep{2005MNRAS.356..654G}. Under the assumption of a Gaussian
core density profile whose amplitude varies as a time-dependent Gaussian, they probed the
chemical evolution at various times and points in the core with a 221 species chemical
code. It was found that the freeze-out in the high-density
regions maintains for some time after the strongest compression has passed until the extinction
drops below a critical value and the molecules are evaporated back into the gas phase.
This result would be consistent with finding a core in the process of
decreasing density and yet showing substantial depletion.

While running a chemistry model for a core with the properties of L1506C and a given
evolutionary sequence for the density structure is possible, comparison with the line
data would also require to run a line transfer model. The parameter space to be covered
will need careful exploration and this advanced effort is beyond the scope of this paper,
but an interesting perspective for future work on this remarkable core.

The presented results and both scenarios are in agreement with L1506C being a pre-stellar core 
in the making as proposed by Pea10: 
despite its turbulent and dense history the low-density core
appears to be gravitational bound and showing infall motions.
But within the picture of grain growth in cores,
the presence of coreshine and depletion might indicate that it is in the
re-making.

\begin{acknowledgements}
We thank David Flower and Wing-Fai Thi for valuable comments, Laurent Pagani
for comments and providing the MAMBO map,
and the referee for suggestions helping to improve the
paper.
JS and MA acknowledge support from the ANR (SEED ANR-11-CHEX-0007-01).
CWO acknowledges support for this work by NASA
through Hubble Fellowship grant No. HST-HF-51294.01-A
awarded by the Space Telescope Science Institute, which is
operated by the Association of Universities for Research in
Astronomy, Inc., for NASA, under contract NAS 5-26555. 
This work is based on observations made with the Spitzer Space Telescope,
which is operated by the Jet Propulsion Laboratory,
California Institute of Technology under a contract with NASA.
\end{acknowledgements}

\bibliographystyle{aa}
\bibliography{L1506C}

\end{document}